\documentclass[prb,twocolumn,preprintnumbers,amsmath,amssymb]{revtex4}
\makeatother
\usepackage[dvips]{graphicx}
\usepackage{amsmath}
\usepackage{amsbsy}
\usepackage{caption}
\usepackage{color}
\DeclareGraphicsExtensions{.pdf,.eps,.png,.jpg,.mps}
\usepackage{gensymb}
\usepackage{float}

\definecolor{textcolor}{cmyk}{0,0,0,1}
\definecolor{magenta}{rgb}{1,0,1}
\definecolor{green}{rgb}{0,1,0}
\definecolor{red}{rgb}{1,0,0}

\usepackage{xcolor}

\begin{document}

\title{
Stable Carbon Monosulfide Nanostructures: Chains Arrays and  Monolayers
}
\author{ T. Alonso-Lanza,$^{1}$  F. Aguilera-Granja,$^{1,2}$  J. W. Gonz\'alez$^{1}$  A. Ayuela$^{1}$}
\affiliation{
$^1$Centro de F\'{\i}sica de Materiales CFM-MPC CSIC-UPV/EHU, Donostia 
International Physics Center (DIPC), Departamento de F\'{\i}sica de Materiales, Fac. de Qu\'{\i}micas, UPV-EHU, 20018 San Sebasti\'an, Spain
\\
$^2$Instituto de F\'{\i}sica, Universidad Aut\'onoma de San Luis de Potos\'{\i}, 78000 San Luis Potos\'{\i} S.L.P., M\'exico
}

\begin{abstract}
Theoretical predictions have lead to the experimental synthesis of new low dimensional layered structures.
  Herein we show for the very first time that compounds of carbon monosulfide exhibit a great variety of layered nanostructures, such as chains arrays, monolayers, and thin films.
We find that the chains arrays are the most stable because they are mainly dimensionality-driven by the $sp^2$ hybridization of sulfur and carbon orbitals. Furthermore, the chains arrays are direct gap semiconductors.
In contrast to thin films, the monolayers are stable at room temperature with a semiconductor phase followed in energy by a metallic phase. Then, we achieve a semiconductor-to-metal phase transition in carbon sulfur monolayers, which can be driven by strain engineering controlling conductivity and carrier mobility.
\end{abstract}
\today
\maketitle

\section{\label{sec:intro}Introduction}

New phenomena in condensed matter physics are based on two-dimensional materials. An example is a single layer of carbon atoms known as graphene \cite{novoselov2004electric}. Graphene characterization shows outstanding structural and electronic properties such as high stiffness or flexibility, and massless Dirac fermions with low resistivity \cite{novoselov2012roadmap,huang2011graphene,geim2009graphene,neto2009electronic,bolotin2008ultrahigh,zhang2005experimental}, which allow for many applications. The main drawback is the absence of band gap in the graphene electronic structure and the small on$/$off ratio\cite{geim2009graphene,schwierz2010graphene}, a band-gap may be opened via different techniques, such as chemical tailoring, inducing strains, and geometric patterning\cite{C6CP04677K,semenoff2012chiral,diaz2017tuning,zhou2007substrate,gonzalez2011transport}.
Although graphene looks promising for the future, research is today focusing on other carbon nanostructured materials and other two-dimensional materials\cite{gogotsi2015not}. 

The possibility to exfoliate materials and the existence of accurate experimental techniques to study single-layer materials\cite{hashimoto2004direct,liu2009open,jin2009deriving} supposed a big boost to study other layered materials.
Layers with other elements in the carbon group have been studied to produce counterparts to graphene, such as silicene, germanene or stanene\cite{vogt2012silicene,lalmi2010epitaxial,aufray2010graphene,balendhran2015elemental}.  Graphene is flat because there is \textit{sp$^2$} hybridization, and these other compounds adopt a buckled structure because they prefer \textit{sp$^3$} hybridization. Other layered nanostructures isoelectronic to graphene are obtained either combining III-V group elements, such as in hexagonal boron nitride\cite{zunger1976optical}.
2D materials research is recently widen to other new materials based on phosphorus, whose isoelectronic counterparts are a central issue nowadays\cite{zhu2015designing,yang2016two,alonso2016array}. 
Black phosphorus is formed by single layers mediated by van der Waals interaction, so that analogously to graphene, few layers are isolated to form phosphorene \cite{liu2014phosphorene,castellanos2014isolation}. Phosphorene has a two dimensional honeycomb puckered structure where each atom is bonded to three neighbors. It displays a direct band gap, tunable attending to the number of layers and the stacking\cite{xia2014rediscovering,dai2014bilayer}.

More and more systems are being proposed and deeply analyzed to obtain 2D materials with improved stability and new properties\cite{BCN}; one of the hottest topic nowadays focuses on phosphorene isoelectronics compounds.

There are two different approaches to obtain phosphorene isoelectronic compounds, by taking elements of the group V or by mixing elements of the groups IV-VI also called group IV monochalcogenides\cite{gomes2015phosphorene,hong2016electronic,fang2016layer,gomes2015enhanced,tuttle2015large,mehboudi2016two}. 
The  phosphorene isoelectronic compounds of the group V are nitrogene, phosphorene, arsenene\cite{kamal2015arsenene,mardanya2016four,zhang2015atomically} and antimonene\cite{zhang2015atomically}, and their binary compounds\cite{yu2016atomically,xiao2016prediction}. They are currently undergoing characterization, for instance, by studying doping\cite{wang2016electronic}, point defects\cite{wang2016electronic,gomes2016vacancies}, and oxygen contamination\cite{de2016effects}. 
The group IV monochalcogenides includes a combination of light elements, such as in silicon monosulfide monolayers and thin films\cite{zhu2015designing,yang2016two,alonso2016array}, and a combination of heavy (Ge, Sn) and light (O, S) atoms \cite{ma2016structure}, such as in silicon telluride\cite{chen2016site} and germanium monosulfide\cite{li2016germanium} monolayers, which exhibit strain-tunable indirect band gap. Nevertheless, none of the today proposed phosphorene isoelectronic compounds is based on carbon. 
In fact, the layered carbon compounds, such as carbon nitride\cite{de2016mechanical} and carbon phosphide\cite{guan2016two,wang2016carbon}, are not isoelectronic to phosphorene.  
Our aim is to design new layered nanostructures isoelectronic to black phosphorus, but still composed of carbon.

We present here for the first time stable carbon monosulfide nanostructures covering chains, monolayers,  and thin films. We have performed phonon and molecular dynamics calculations to ensure the stability of those structures at room temperature. The carbon monosulfide nanostructures vary from metallics in thin films to semiconductors in chains and most monolayers. The chains and thin films are more stable than the monolayers because sulfur atoms have two bonds instead of three (or four) for the monolayers \cite{alonso2016array}. The large variety of nanostructures that we explored with either metallic or semiconductor character makes this material an interesting compound for several applications based on electronic transport within semiconductors devices.

\section{\label{sec:comp}Computational details}

We use the SIESTA (Spanish Initiative for Electronic Simulations with Thousands of Atoms) package to carry out density functional theory (DFT) calculations for carbon monosulfide nanostructures. For the exchange and correlation potentials we used the generalized gradient approximation (GGA) in the form of Perdew-Burke-Ernzenhof\cite{perdew1996generalized}.
We fully relax atomic positions and unit cells until forces are below 0.006 eV/\AA{}. 
The used unit cells have large vectors about 24.5~\AA{} in the perpendicular direction to the layers in order to avoid interaction with images.
An electronic temperature of 25 meV and a meshcutoff of 250 Ry are used. The sampling of the Brillouin zone has a fine grid of 20$\times$20$\times$1 k points. 
The atomic cores are described by nonlocal norm-conserving Troullier-Martins\cite{troullier1991efficient} pseudopotentials factorized in the Kleynman-Bylander form.
Orbitals are developed in basis sets on each atom, and the basis size is double zeta plus polarization orbitals. 
Details on the used pseudopotentials and basis were described previously \cite{alonso2015chemical,C6CP04677K,alonso2016array}. Anyhow our main results are checked by repeating calculations with the VASP code using the projected augmented wave method (PAW) and within the PBE formalism for the exchange and correlation \cite{blochl1994projector,kresse1999ultrasoft}, and the {\sc Quantum ESPRESSO} method with equivalent parameters\cite{QE-2009}. 
We estimate the van der Waals contribution to the total energy for a two dimensional array with well separated chains using the implementation within the VASP code\cite{klimevs2011van,roman2009efficient} of the vdW-DF\cite{dion2004van} functional.
Last but not least, we even confirm the stability of such highly stable obtained structures by carrying out molecular dynamics calculations within the SIESTA code using the Nose thermostat at room temperature. We employed a Nose mass\cite{souza2012ab,zanolli2009defective,hobi2010formation} of 10 Ry fs$^{2}$, and a time step of 1 fs. We have chosen 3000 as final time step and the relaxation time to reach target temperature was 2500 fs.
Phonon dispersion curves are obtained using the {\sc Quantum ESPRESSO} code, with details  included in the Supplemental Information.

\section{\label{sec:results}Results and discussion}

We study three different types of carbon monosulfide nanostructures, monolayers, thin films, and chains arrays, following the order from lower to higher stability. Molecular dynamics simulations show that thin films are not stable at room temperature. Comments and analysis of thin film structures are therefore presented in the Supplemental Information, together with further information on monolayers and chains.

\begin{figure*}[thpb]
      \centering
\includegraphics[clip,width=\textwidth,angle=0,clip]{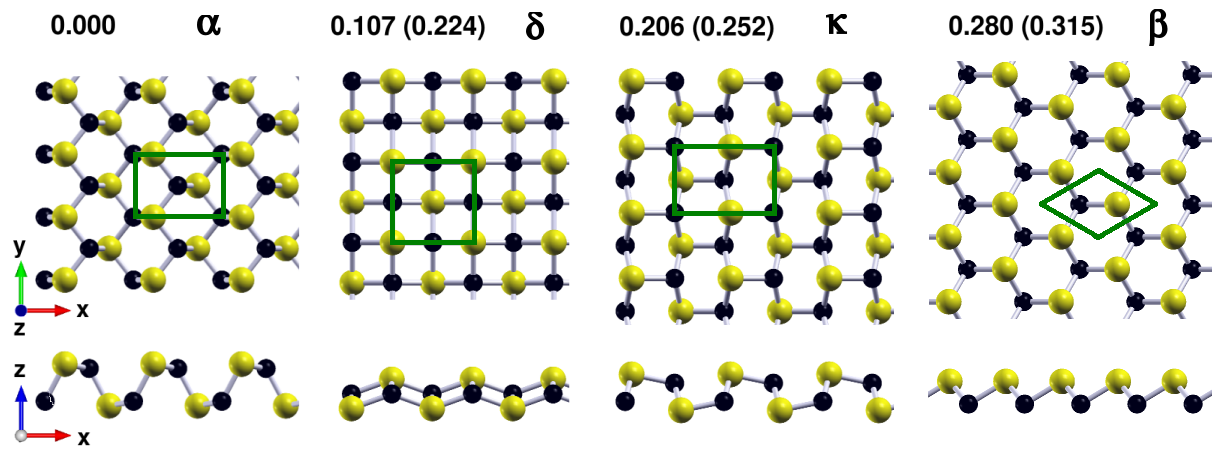}
\caption{\label{estructuras} 
Geometries of the four carbon monosulfide monolayers in order of decreasing stability. Carbon (sulfur) atoms are represented by black (yellow) spheres through the article. The unit cell is shown in green for each case. Differences in the energy per atom using the SIESTA and VASP methods are shown on top and in parentheses for each structure, respectively. The energy differences are calculated relative to the energy per atom in the $\alpha$ structure.}
\end{figure*}


\subsection{\label{sec:monolayers} Monolayers}

\begin{figure*}[thpb]
      \centering
\includegraphics[clip,width=\textwidth,angle=0,clip]{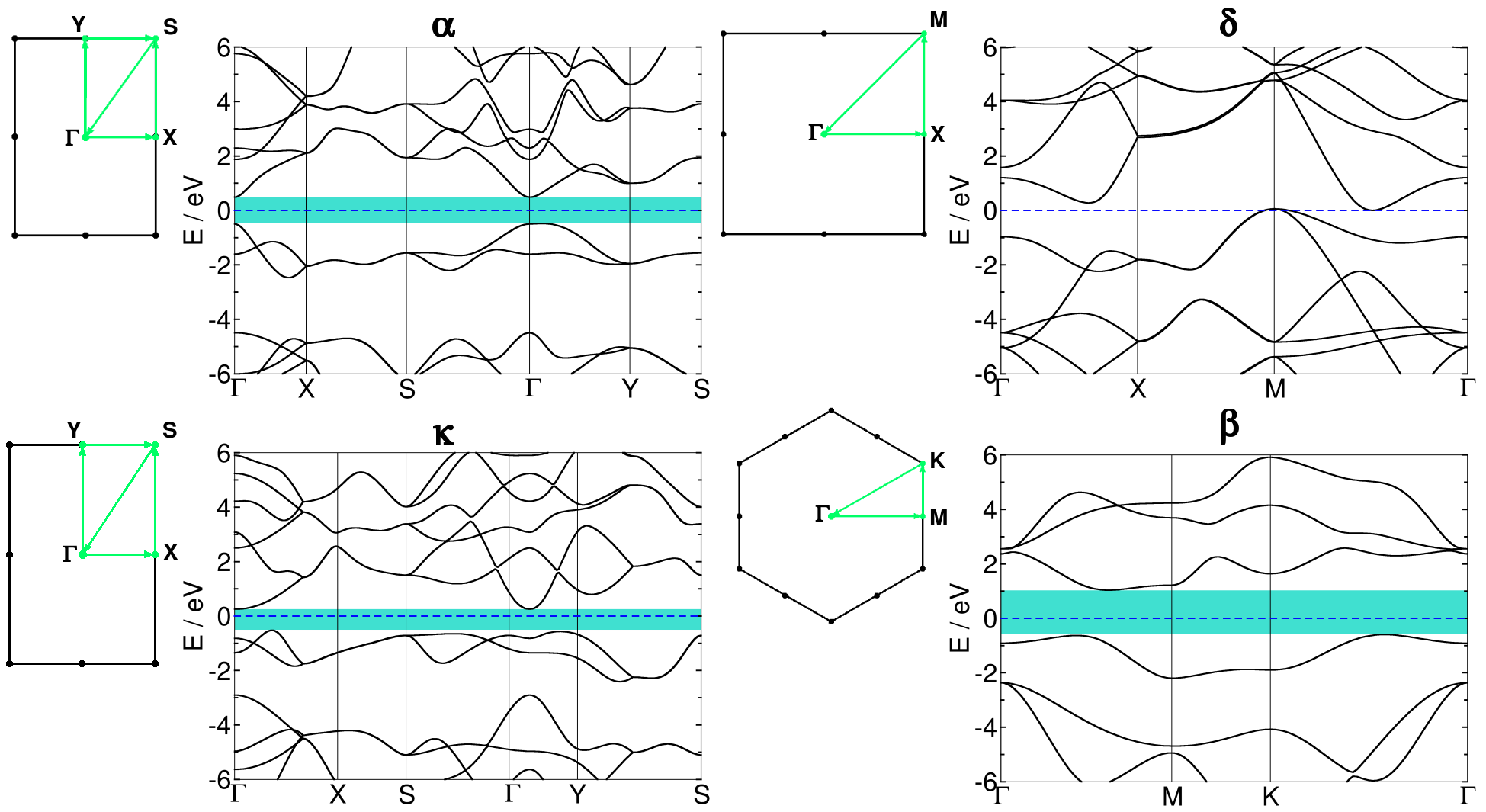}
\caption{\label{bandas} 
Band structures of the four carbon monosulfide monolayers. Fermi energies are denoted by horizontal dashed lines. Band gaps over the whole Brillouin zone are shown by shaded rectangles. }
\end{figure*}

We explore a wide variety of structures for carbon monosulfide monolayers. We next present the geometry and electronic structure of the four most stable structures, and finally consider the ground state in more detail.
Figure \ref{estructuras} shows four monolayers of carbon monosulfide after structural relaxation including their relative energies. The most stable $\alpha$ monolayer  presents a similar structure to that of a single layer of black phosphorus. This structure has also been reported recently for group IV monochalcogenides\cite{gomes2015phosphorene,zhu2015designing}, and group V semiconductors such as phosphorous-nitrogen or arsenic-nitrogen \cite{zhang2016tinselenidene,yu2016atomically,xiao2016prediction}. The results obtained with VASP method yield energy differences slightly larger, although the stability order is the same. Second in energy we find the $\delta$ monolayer that is a regular structure composed of squares looked from above. A side view reveals two different heights for sulfur atoms. Next there is a similar structure, the $\kappa$ structure, with all the carbon atoms shifted towards one side producing long and short distances with the sulfur atoms. 
When the distortion is maximized the $\alpha$ structure is reached. The  $\kappa$ layer seems an intermediate structure between the $\alpha$ and $\delta$ monolayers, so that the three monolayer monosulfide-carbon structures seems related, to be discussed at length in next section \ref{sec:strain}. 
Phonon dispersions, shown in Supplemental Information, indicate that the $\alpha$ and $\delta$ structures are really stable, and the $\kappa$ structure is unstable because there are negative frequencies. Furthermore, molecular dynamics simulations show that the $\alpha$ and $\delta$ structures are stable even at room temperature. 
Last we find the  $\beta$ structure which from a top view it displays a hexagonal pattern, and in a side view it has a buckled structure, reported for other phosphorene isoelectronic compounds\cite{zhu2015designing}. The $\beta$ structure is similar to that of blue phosphorene\cite{zhang2016epitaxial,zhu2014semiconducting}.
The phonon dispersion curves show that the carbon monosulfide $\beta$ phase is unstable.

\begin{figure*}[thpb]
      \centering
\includegraphics[clip,width=\textwidth,angle=0,clip]{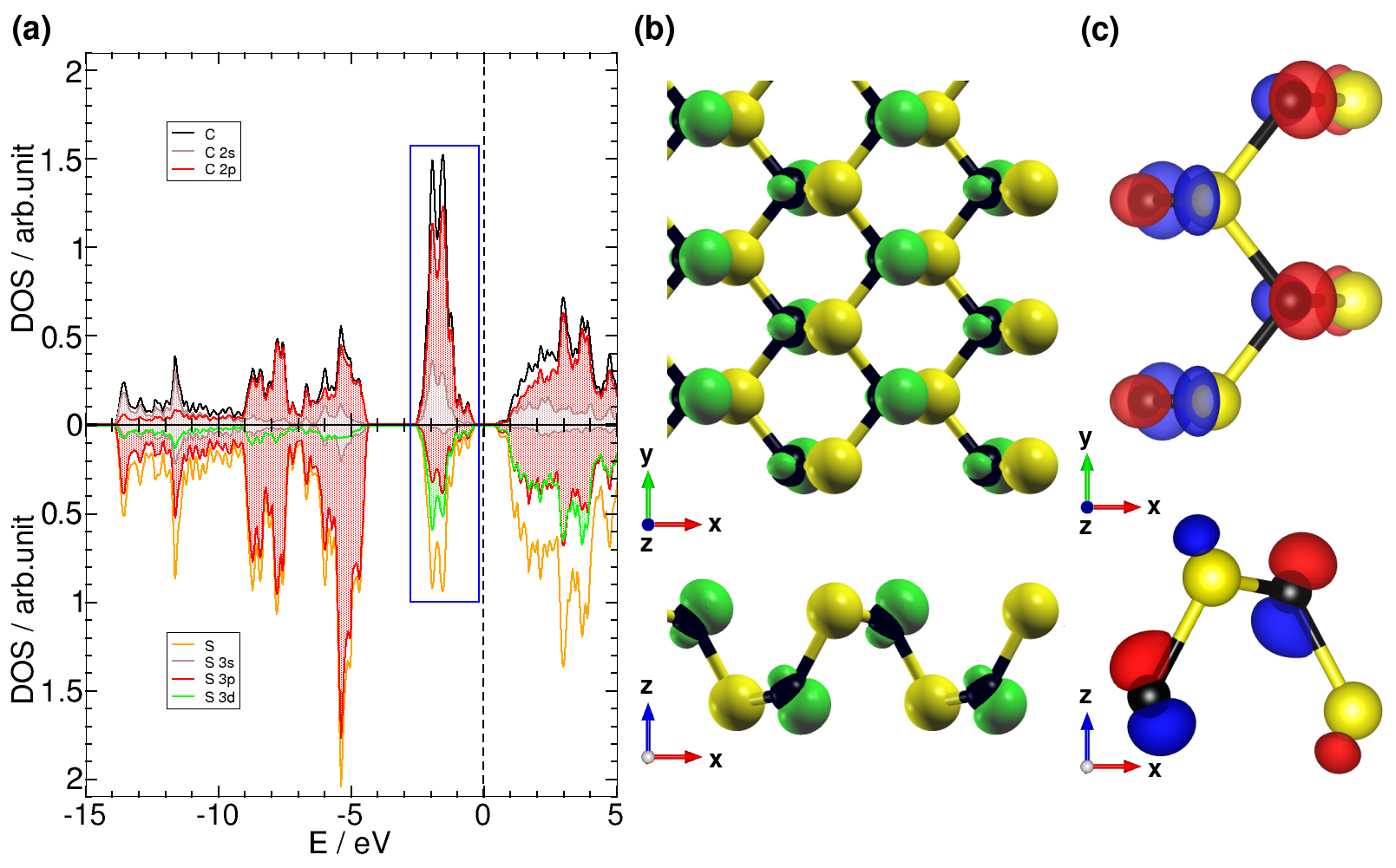}
\caption{\label{alpha} 
Electronic properties of the ground $\alpha$ structure. (a) Density of states projected on carbon (up panel) and sulfur (down panel). (b) Spatial localized density of states for the energy peak marked up with a blue rectangle in (a). (c) Highest occupied orbital of the valence band at the $\Gamma$ point.}
\end{figure*}

To gain insight on the electronic structure of the carbon monosulfide monolayers we consider the band structures shown in Fig. \ref{bandas}. On the one hand, three of the monolayers, labeled as $\alpha$, $\kappa$ and $\beta$, are semiconductors with indirect band gaps of 0.97, 0.77 and 1.64 eV, respectively. For the $\alpha$ and $\kappa$ layers, the minimum of the conduction band is located at the gamma point. For the $\alpha$ case, the highest occupied valence band is well separated from lower valence bands with a pseudogap of 2 eV. The $\kappa$ structure has still a pseudogap opening in the valence band, although the value is reduced to  0.5 eV. 
On the other hand, the $\delta$ structure is metallic.

The bands of the $\alpha$, $\delta$ and $\kappa$ structures appear similar because they share common structural features as commented above. For the $\beta$ monolayer the bands are degenerated throughout the Brillouin zone due to the lattice symmetry. Most of the features of the bands are the same for the $\alpha$ and $\kappa$ cases; for instance there are two occupied bands just below the Fermi energy, extended along all Brillouin zone, and well separated from the deeper states.  However, the band structure of the $\delta$ case differs because one of the two occupied bands close to the Fermi energy becomes broad and metallized, going down to deep energies following  a parabolic shape. The two occupied narrow bands for the $\alpha$ and $\kappa$ cases are localized within a interval of 2 eV, while the bands for the $\delta$ case become broad expanding over more than 6 eV. We find two unoccupied conduction bands that mirror the behavior of two nearest valence bands with respect to the Fermi energy.  There are even bands for the $\delta$ structure crossing the Fermi energy, resulting in a conducting monolayer.

We furthermore study the charge transfer between carbon and sulfur atoms using the Mulliken scheme.  The charge values are 0.81, 0.92, 0.65 and 0.66 electrons for the $\alpha$, $\delta$, $\kappa$ and $\beta$ monolayers, respectively. We find that the charge transfer of almost one electron in some cases is large and goes from carbon to sulfur atoms. This result is surprising because carbon and sulfur have almost the same Pauling electronegativity. It seems that the larger size of sulfur atoms with respect to carbon atoms must be taken into account in order to explain the charge gain by sulfur atoms.

\begin{figure*}[thpb]
      \centering
\includegraphics[clip,width=\textwidth,angle=0,clip]{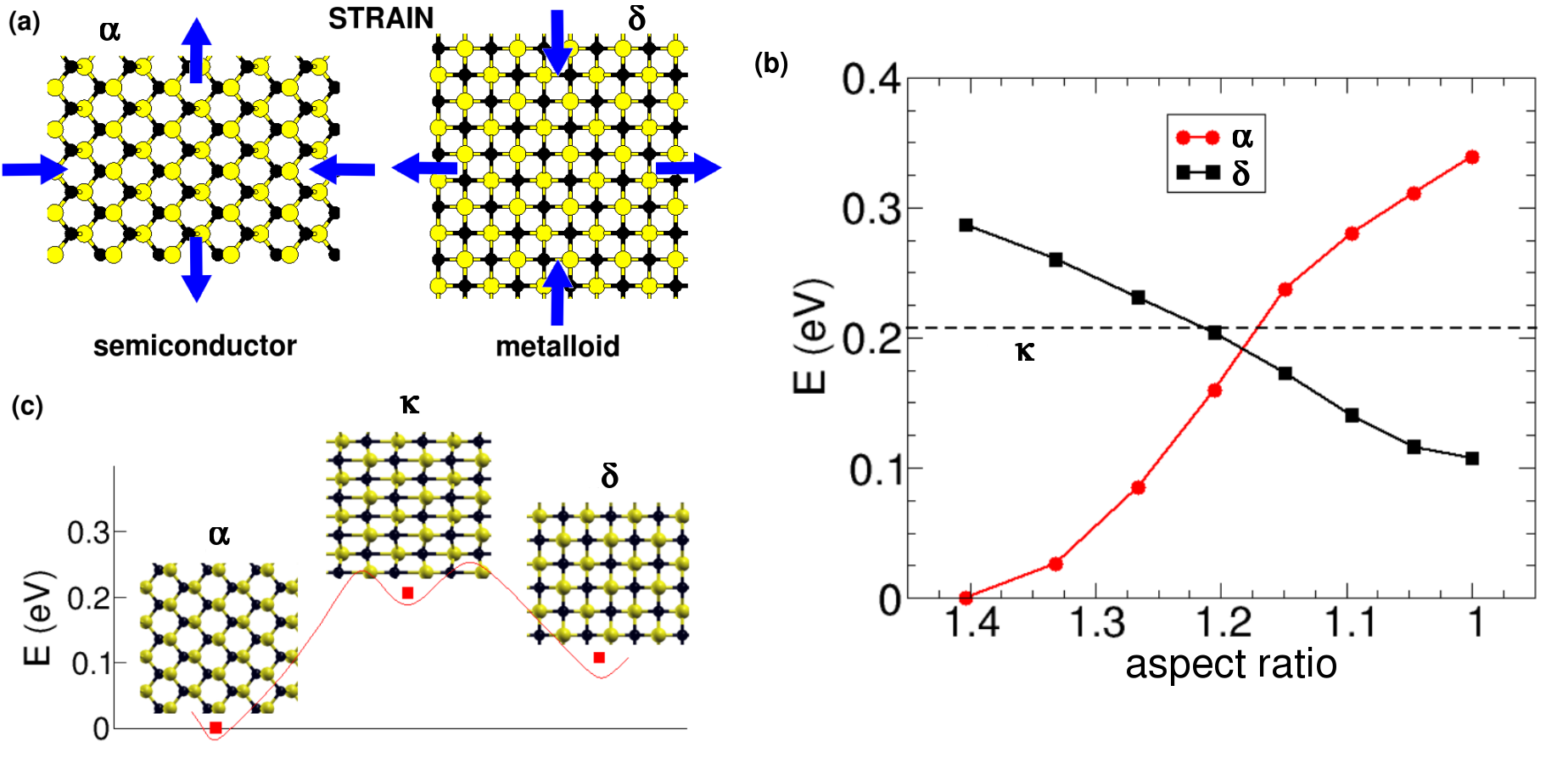}
\caption{\label{strain} 
Stretching the $\alpha$ and $\delta$ monolayer as shown in panel (a) yields an increasing energy as shown in panel (b), suggesting a barrier between those structures. (c) Relative position in energy of the three most stable monolayers sketching a possible energy  profile.}
\end{figure*}

\subsubsection{\label{sec:alpha} $\alpha$ monolayer}

We focus on the ground state monolayer given by the $\alpha$ structure. 
Each carbon atom is bonded to three sulfur atoms and vice versa. There are two bonds of 1.82~\AA{} and one of 1.90~\AA{}. The angles between the bonds from a carbon atom are 113$\degree$ in two cases and 103$\degree$ in the other. 
It seems that carbon atoms present a sp$^3$-like hybridization.

We discuss on the relative contribution between carbon and sulfur atoms to discuss stability. The projected density of states is shown in Fig. \ref{alpha}(a). The total states projected on sulfur contribute more than those on carbon because sulfur has a larger number of electrons than carbon. Nevertheless, the occupied peak close to the Fermi energy has a larger carbon contribution than sulfur. This particular peak is related to the isolated valence band, discussed above when commenting Fig. \ref{bandas}.  The density of states (LDOS) in real space within the energy range of 2 eV is included in panel (b). Most of the states are localized around carbon atoms with orbital lobes suggesting a large $p$-type contribution. The peak LDOS globally looks like a dangling bond of the $sp^3$ hybridization.
However, the wave function for the highest occupied orbital in $\Gamma$ point, as shown in Fig. \ref{alpha}c, reveals a $p_z$-like orbital over the carbon atoms, to be held responsible for the broadening of a band close to Fermi energy upon distortions. 
The projected sulfur LDOS shown in Fig. \ref{alpha}(a) has a non negligible $d$ contribution, especially for the highest energy occupied peak. Therefore, $spd$ hybridization on sulfur atoms seems crucial to describe the valence band close to the Fermi energy\cite{alonso2016array}.

We then compare carbon-sulfur and silicon-sulfur \citep{zhu2015designing,alonso2016array} compounds, both being isoelectronic to phosphorene, in the $\alpha$ structure. There are two well differentiated roles depending of the atom size, as shown in the $\alpha$ structure of Fig. \ref{estructuras}. The smaller atom, either carbon for CS or sulfur for SiS, is the inner one, and the larger atom is the outer one. 
The electronic structure produces similar trends taking into account that the inner atom contributes with $p_z$ orbitals to the highest occupied orbital at the $\Gamma$ point. Figure \ref{alpha}(c) shows a $ p_z$-type orbital for carbon atoms in carbon-sulfur compounds. The main contribution to the frontier orbitals in silicon-sulfur compound comes from the $p_z$ orbitals in sulfur\citep{zhu2015designing}. 
 These bands for the outer atom project on a hybridized dangling bond. It has $pd$ hybridization for sulfur in CS, as shown by the projected density of states in Fig. \ref{alpha}(a), and it has $spd$-like hybridization for the frontier orbitals of silicon in SiS \citep{zhu2015designing,alonso2016array}. It is noteworthy that sulfur atoms can adopt the two roles depending on the counterpart element involved in the isoelectronic compound to phosphorene.

\begin{figure*}[thpb]
      \centering
\includegraphics[clip,width=\textwidth,angle=0,clip]{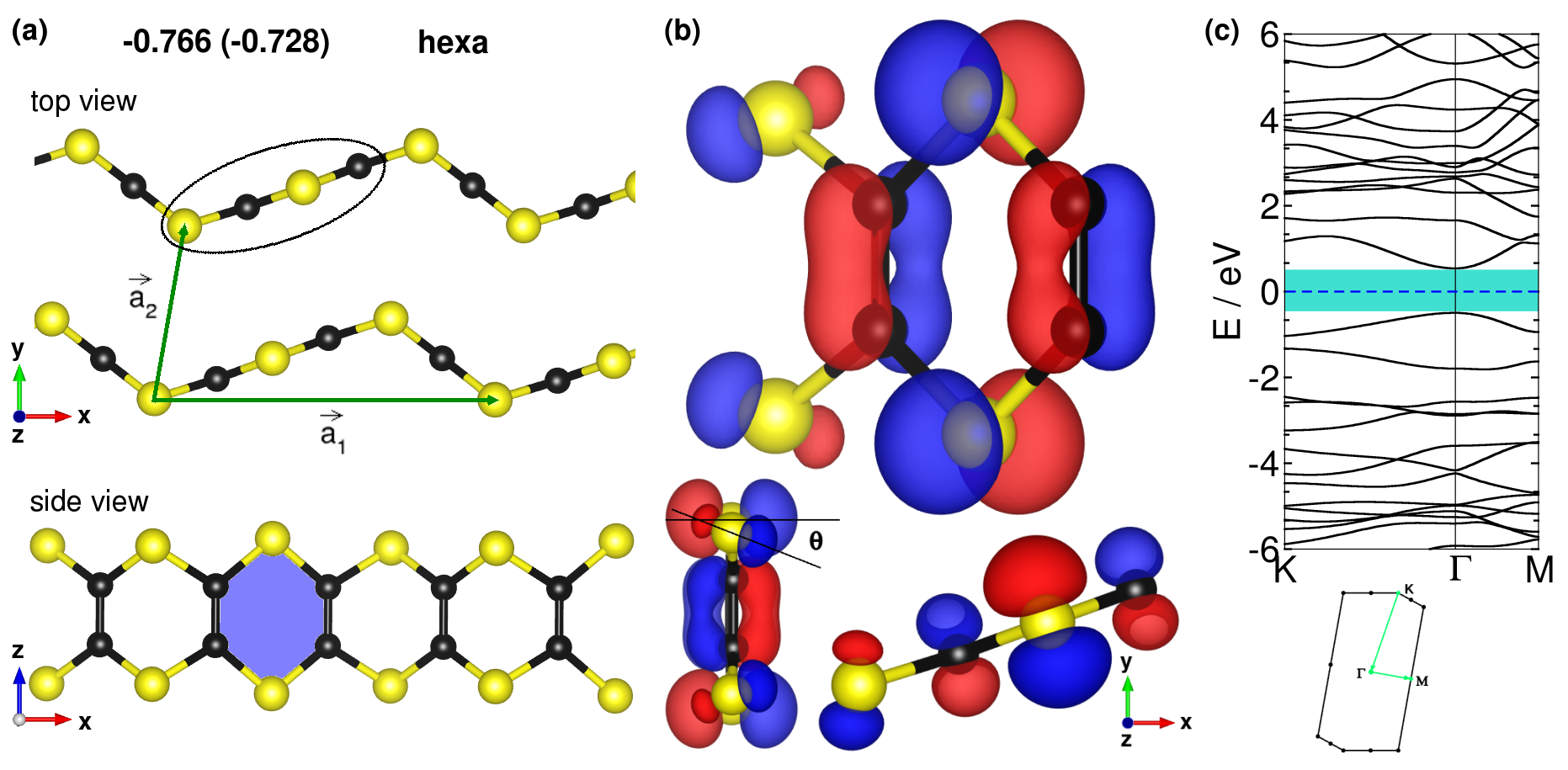}
\caption{\label{chains} 
(a) Geometry of the \textit{hexa} chains arrays, with the hexagon motif of the structure highlighted in blue color. The unit cell to compute the band structures are marked by vectors. The energies per atom obtained using SIESTA and VASP codes are given above referred to the ground state $\alpha$ monolayer. (b) Highest occupied molecular orbital (HOMO) at the $\Gamma$ point for the \textit{hexa} chains in the region marked with a black ellipse in panel (a). (c) Band structure of the \textit{hexa} chains.}
\end{figure*}

\subsubsection{\label{sec:strain}Strain-driven transition between two stable monolayer phases}

We try to shed light over the striking metallization of the $\delta$ monolayer. A distortion of the $\alpha$ atomic positions could end in the $\delta$ structure, so that  the monolayer could undergo a semiconductor-metal transition driven by a strain field. The $\delta$ structure is isotropic and similar to a rippled square lattice.
Sulfur atoms show $spd$ hybridization as denoted by the projected density-of- states, similarly to the $\alpha$ structure. A detailed analysis of the localized density of states shows that the Fermi energy states are $p$-type on carbon, which could be seen as a square lattice of $p_z$ orbitals. Details on the geometry and the projected and localized density-of-states of the $\delta$ phase are found in the Supplemental Information.

We follow the transition between the $\alpha$ and $\delta$ monolayers by changing aspect ratio, as shown schematically in Fig. \ref{strain}(a). To this aim we modify the unit cell shape step-by-step and transform the square cell into a rectangular one. The unit cell following the transformation is characterized by the aspect ratio between length and width, which is 1 for the $\delta$ phase and ~1.4 for the $\alpha$ phase. For each calculation the unit cell remains fixed and the inner atomic positions are relaxed. We expected that the $\delta$ monolayer changing aspect ratio would transform into the $\alpha$ monolayer by relaxing the inner atomic positions. However, we find that the inner atomic structure is retained when deforming the unit cell, and the energy increases continuously with the aspect ratio, as shown in Fig. \ref{strain}(b). We also find a similar behavior when deforming the $\alpha$ monolayer. The two curves of energy versus aspect ratio cross at a point with a barrier of about 0.2 eV between the monolayers, a fact sufficient to confirm the stability for the $\delta$ structure.
The energy of the $\kappa$ monolayer, included in Fig. \ref{strain}(b), is very close to the crossing point, which supports the idea of $\kappa$ being a metastable phase between $\alpha$ and $\delta$ monolayers. Figure \ref{strain}(c) sketches a possible energy profile following this picture.

\subsection{\label{sec:chains}Chains arrays}

We find other kind of structures for carbon monosulfide with chains arrays. The previously studied $\gamma$ SiS monolayer\cite{alonso2016array}  is fully relaxed by substituting silicon with carbon atoms. A layer with chains of carbon monosulfide is then stabilized by more than half electronvolt per atom when compared to the most stable $\alpha$ monolayer. 
We obtained two different chains arrays with hexagonal and pentagonal motifs. The \textit{hexa} chains are more stable than the \textit{penta} chains by more than 0.25 eV per atom, so we focus on the \textit{hexa} chains below. The \textit{penta} chains are commented in the Supplemental Information. 
 The structure with hexa-motif chains, shown in Fig. \ref{chains}(a), is 0.766 eV per atom more stable than the ground state $\alpha$ monolayer, which results in a large rise of stability. 
The distance between the closest atoms from two different chains is larger than 3.7~\AA{}, which means that interaction between adjacent chains is mediated by dispersion forces. Thus, the bonds between chains are much weaker than those established within a single chain. \footnote{Although the interaction between chains by dispersion forces is beyond the actual scope of the paper, we have tested the \textit{hexa} chains including van der Waals interaction using the VASP code, obtaining that the stability is enhanced by approximately 70 meV per atom.}

\begin{figure*}[thpb]
      \centering
\includegraphics[clip,width=\textwidth,angle=0,clip]{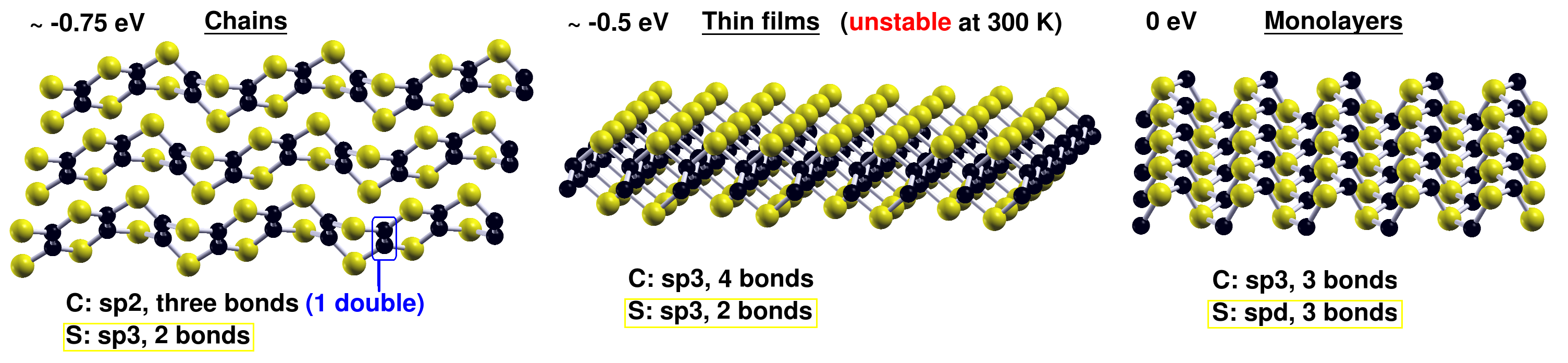}
\caption{\label{resumen} 
Comparison of energetic stability, number of bonds, and hybridizations for carbon monosulfide layered nanostructures. Note that the most stable structures are the chains arrays.}
\end{figure*}

We investigate in more detail the bonds between atoms in the chains. The top view in Fig. \ref{chains} (a) shows that due to kinks, the \textit{hexa} chains combines long and short stretches. The kinked chains are more stable by an 70 meV per atom than the straight chains. We check the structural stability of this structure with chains using molecular dynamics simulation for a single chain at room temperature. 
Carbon atoms form covalent bonds with sulfur atoms. Each carbon atom is bonded to two sulfur atoms with two bond lengths of about 1.82~\AA{} and to another carbon atom with a bond length of 1.38~\AA{}. The short bond length between carbon atoms indicates the existence of a double bond with a $\sigma$ bond and a $\pi$ bond. The $\pi$ bond is spatially shown by the highest occupied molecular orbital at the $\Gamma$ point in Fig. \ref{chains}(b), formed by two $p$-type orbitals from each pair of carbon atoms. 
Concerning the angles between the three bonds, there are two different carbon atoms depending on whether they are in a short stretch or in a long stretch; however, for all the cases the angles are close to 120$\degree$. 
The fact that the three bonds are planar and form angles close to 120$\degree$ is assigned to a $sp^2$ hybridization. The remaining $p_z$ orbital is used to form the double $\pi$ bond with the neighboring carbon atom.
Focusing now on the sulfur atoms, each atom establishes two bonds with carbon. The chain adopts a zigzag pattern with two consecutive sulfur atoms in kinks followed by other in a straight stretch.
The two types of sulfur atoms seem to have the same hybridization. They contribute with $p$-orbitals to the highest occupied orbital at the $\Gamma$ point, as shown in Fig. \ref{chains}(b). However, the kinks effect is to tilt the $p_z$ orbitals by an angle $\theta$ of approximately 25$\degree$, as presented in Fig. \ref{chains}(b).

The analysis of the projected density of states for the \textit{hexa} chains array, attached in the Supplemental Information, shows that close to the Fermi energy there are mainly $p$-type contributions, in contrast to the  $\alpha$ and $\delta$ monolayers.
The charge reorganization under the Mulliken scheme results in a net transfer of approximately 0.6 electrons from each carbon atom towards the sulfur atoms. Furthermore, we analyze the band structure shown in Fig. \ref{chains}(c) along two directions, parallel and perpendicular to the chains. We find that the \textit{hexa} chains array is semiconductor with a direct band gap of 1.03 eV at the $\Gamma$ point, a fact that could be interesting for electronic applications.

\section{Final remarks}

Finally, we comment on the stability order between the different layered nanostructures of carbon monosulfide we have studied this far.  We find chains, thin films and monolayers in order of increasing energy. The energy is explained by hybridization and the numbers of bonds for carbon and sulfur atoms in each structure, as summarized in Fig. \ref{resumen}.
The  $\alpha$ monolayer, taken as reference,  presents a $sp^3$ hybridization for carbon atoms with three bonds and a dangling bond, and a $spd$ hybridization for sulfur atoms with three bonds.
The thin films have lower energies per atom because the carbon atoms establish a fourth bond and the sulfur atoms lost a bond. Note that sulfur atoms prefer to establish two bonds to fulfill the octet rule\cite{yang2016two,alonso2016array}. The thin films are unstable at room temperature in molecular simulations. 
Last but not least, the \textit{hexa} chains array improves energy and stability with respect to thin films. 
The gain in stability for chains is ascribed to the carbon preference for the $sp^2$ hybridization with two single bonds and a double bond over the fourfold coordination of the $sp^3$ hybridization in thin films. 

In summary, we present brand new stable nanostructures for carbon monosulfide, which include stable chains arrays and monolayers. On the one hand, the chains arrays, which are more stable than monolayers, present hexagonal and pentagonal patterns with carbon and sulfur atoms displaying $sp^2$ hybridizations. These results pave the way so that this type of nanostructures with carbon monosulfide chains can be synthesized  as freestanding chains, but probably better grown supported on some substrates as chains arrays.
On the other hand, the monolayers show $sp^3$ and $spd$ hybridizations for carbon and sulfur atoms, respectively. The ground state $\alpha$ monolayer  has the structure of a single layer of black phosphorus with an indirect band gap of approximately 1 eV. We find that strain can induce a transition between the most stable $\alpha$ and $\delta$  monolayers. Because the $\alpha$ monolayer is semiconductor and the $\delta$  monolayer is metallic, the $\alpha$ to $\delta$ transition could just allow to control conductivity by applying strain, an effect that looks promising for the design of new electronic devices.

\begin{acknowledgments}
This work has been partially supported by the Projects FIS2013-48286-C02-01-P and FIS2016-76617-P of the Spanish Ministry of Economy and Competitiveness MINECO, the Basque Government under the ELKARTEK project(SUPER), and the University of the Basque Country  (Grant No. IT-756-13). TA-L acknowledge the grant of the MPC Material Physics Center - San Sebasti\'an. FA-G acknowledge the DIPC for their generous hospitality. Authors also acknowledge the DIPC computer center.
\end{acknowledgments}


\clearpage

\section*{\label{sec:appendix} Supplemental Information}

\begin{figure*}[thpb]
      \centering
\includegraphics[clip,width=\textwidth,angle=0,clip]{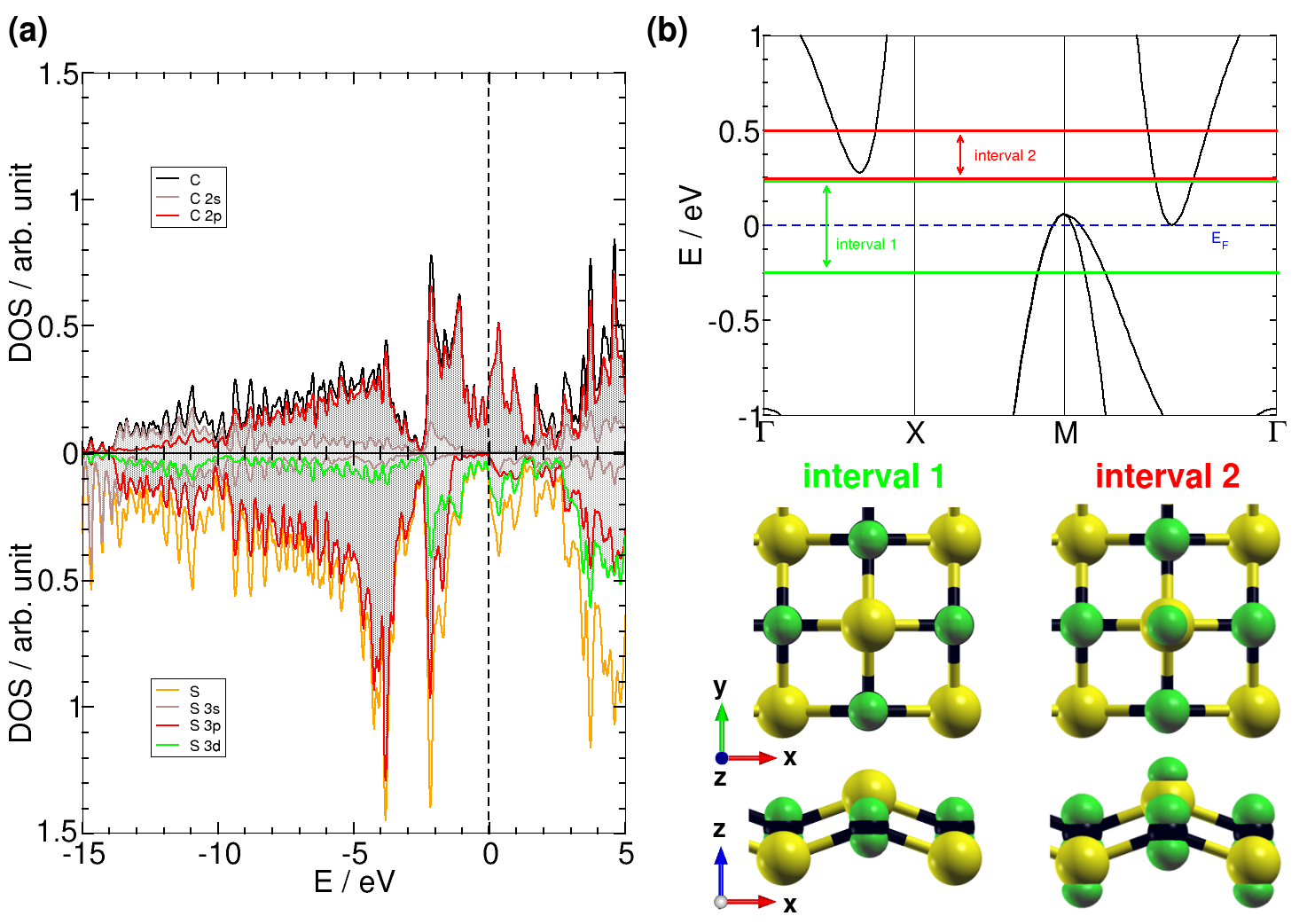}
\caption{\label{square} 
Electronic properties of the $\delta$ monolayer. (a) Density of states projected on carbon (top) and sulfur (bottom). (b) Spatial localized density of states for the two energy intervals marked in the band structure around the Fermi energy.}
\end{figure*}

\subsection{Monolayers}

\subsubsection{Monolayers parameters}

\begin{table}[h]
\small
\caption{Lattice-vectors, first-neighbor distances, and thickness $\Delta_z$ (in \AA{}) for the four carbon monosulfide monolayers. Band gap (in eV). }
\label{table1}
\begin{tabular}{cccccc}
\hline
\hline
 & $\vec{a}$ &  $\vec{b}$  &   d$_{C-S}$ & $\Delta_z$ & gap\\
\hline
\hline
$\alpha$ &  (4.00, 0)   & (0, 2.85)  & 1.90, 1.82 ($\times$2)           & 1.89     & 0.97  \\
$\delta$ & (3.65, 0)    & (0, 3.65)  & 1.93 ($\times$4)                     &  1.31    & 0.77  \\
$\kappa$ &  (4.42, 0)   &  (0, 2.94) & 1.93 ($\times$2), 1.96, 2.53  &  1.61    &    -     \\
$\beta$  & (2.55, 1.47)    & (2.55, -1.47)  & 1.99 ($\times$3)                     &  1.03    & 1.64  \\
\hline
\hline
  \end{tabular}
\end{table}

\subsubsection{$\delta$ monolayer}
We further study the striking metallization of the $\delta$ monolayer. A distortion of the $\alpha$ atomic positions could end in the $\delta$ structure, meaning that by engineering with a strain field the monolayer could undergo a semiconductor-metal transition. The $\delta$ structure is isotropic and similar to a rippled square lattice.  All the carbon-sulfur bonds are equal with a distance of 1.93~\AA{}. 
The consecutive carbon bonds have four angles of 96$\degree$, the opposite bonds have two angles of 141$\degree$. The angles between the sulfur bonds are four of 84$\degree$.
The right angles for sulfur atoms are related to $d$ orbitals taken part in $spd$ hybridization, as shown using the density of states in Fig. \ref{square}(a). However, the occupied states close to the Fermi energy present a major contribution from the states p of carbon and a small contribution with d-character from sulfur. 
The spatial localized density of states (LDOS) for two energy intervals is shown in Fig. \ref{square} (b).
Around to the Fermi energy, the states in interval 1 are localized on carbon atoms, like a square lattice of $p_z$ orbitals explaining metallic behavior. They  become metallic along the $\Gamma$M direction in reciprocal space, equivalent to the $(110)$ direction in real space among carbon atoms. Along the $\Gamma$X direction, there are also empty states very close to the Fermi energy, but with a small gap of 0.22 eV. These states plotted in interval 2 are now including lobules on sulfur towards the outer part of the $\delta$ monolayer.

\subsection{Chains arrays}

\subsubsection{Density of states for the \textit{hexa} chains}

\begin{figure*}[thpb]
      \centering
\includegraphics[clip,width=\textwidth,angle=0,clip]{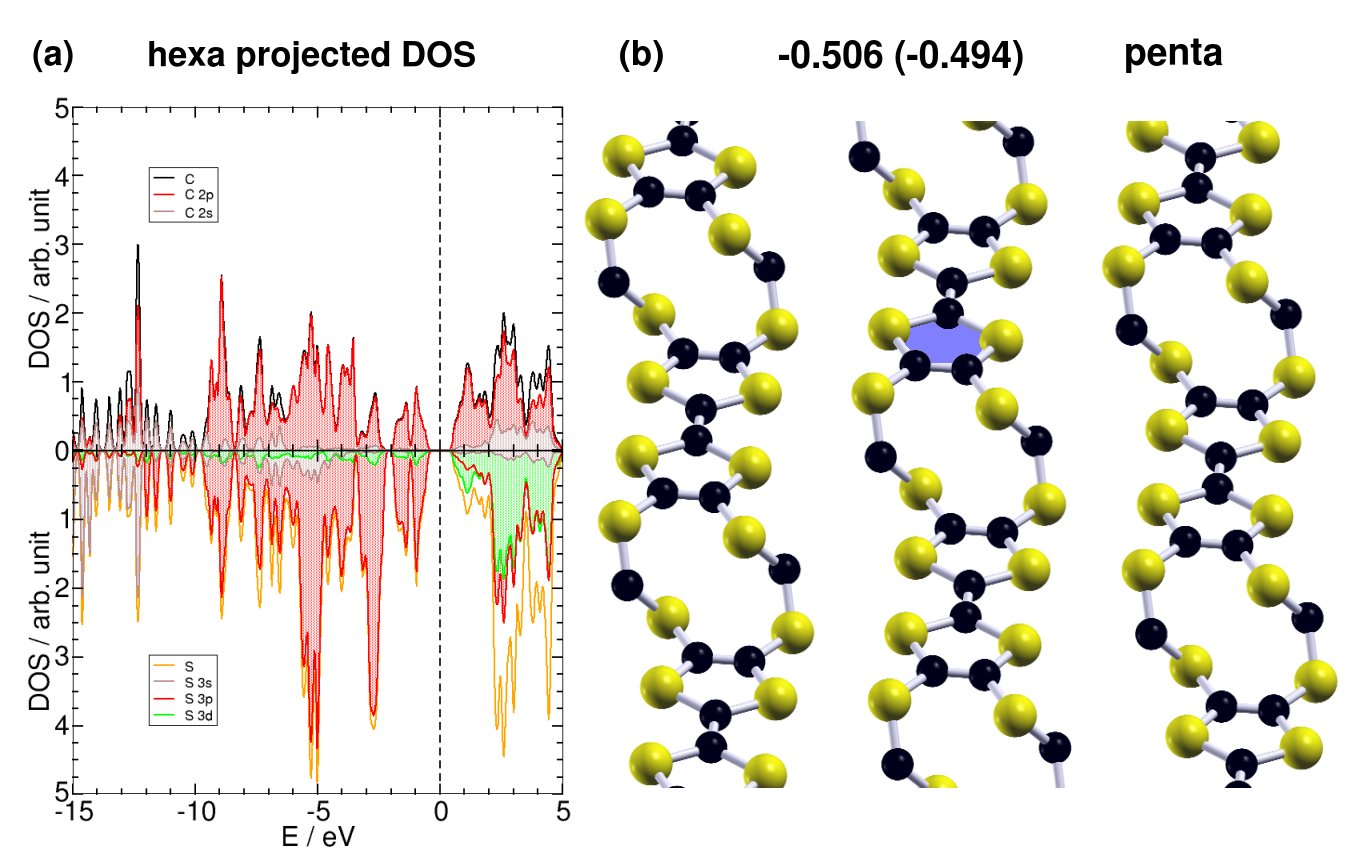}
\caption{\label{hexa} 
(a) Density of states projected on carbon (up) and sulfur (down) for the \textit{hexa} chains array. (b) Geometry and energy of the \textit{penta} chains array. Energy origin is relative to the $\alpha$ monolayer of Fig. \ref{estructuras}.}
\end{figure*}

Figure \ref{hexa}(a) shows the projected density of states resolved per atom site. We find that the $s$ and $p$ orbitals are hybridized at deep energies. Close to the Fermi energy all the states come from the $p$ orbitals centered in carbon and sulfur atoms.  The $d$ orbitals for chains arrays give just residual contributions.

\subsubsection{\textit{Penta} chains arrays}

Because the carbon-carbon bond seems the cause of stability we next investigate structures in which the $C_2$ unit has rotated, as in Stone-Wales rotations. Figure \ref{hexa}(b) shows a \textit{penta} array with chains interacting through their edges. The presence of pentagons motivates the structure name.  The distances between carbon and sulfur atoms between chains are larger than 3.5~\AA{}.  Therefore, there are few atoms interacting through dispersion forces, which means that van der Waals contribution would be lower than for the \textit{hexa} motif.
Focusing on a single \textit{penta} chain, there are two remarkable facts that explain the loss of stability with respect to the \textit{hexa} chains. First, the \textit{penta} motif is similar to the \textit{hexa} motif with a carbon-carbon double bond rotated 90$\degree$, so that instead of having two hexagons we have the two less stable pentagons. The double bonds are preserved with typical bond lengths between carbon atoms of 1.39~\AA{} in most of the cases, and 1.45~\AA{} close to a terrace. The carbon-sulfur bond length is around 1.8~\AA{}. Second, the individual chains have steps each couple of two pentagons linked by a carbon double bond. The angles of carbon bonds are close to 120$\degree$, a characteristic of $sp^2$-like hybridization. 
The two bonds of sulfur atoms have angles about 95$\degree$, except for the sulfur atoms close to terraces where the angle values  ( e.g. 99$\degree$ and 115$\degree$) are slightly larger.
The charge transfer goes from carbon to sulfur with a value larger than 0.5 electrons. Thus, the \textit{penta} array with chains presents hybridizations and a bonding scheme similar as those for the \textit{hexa} motif. It seems that having arrays with chains formed by carbon molecules separated by sulfur atoms is crucial for having high stability with respect to monolayers.

\subsection{Thin films}

\begin{figure*}[h]
      \centering
\includegraphics[clip,width=\textwidth,angle=0,clip]{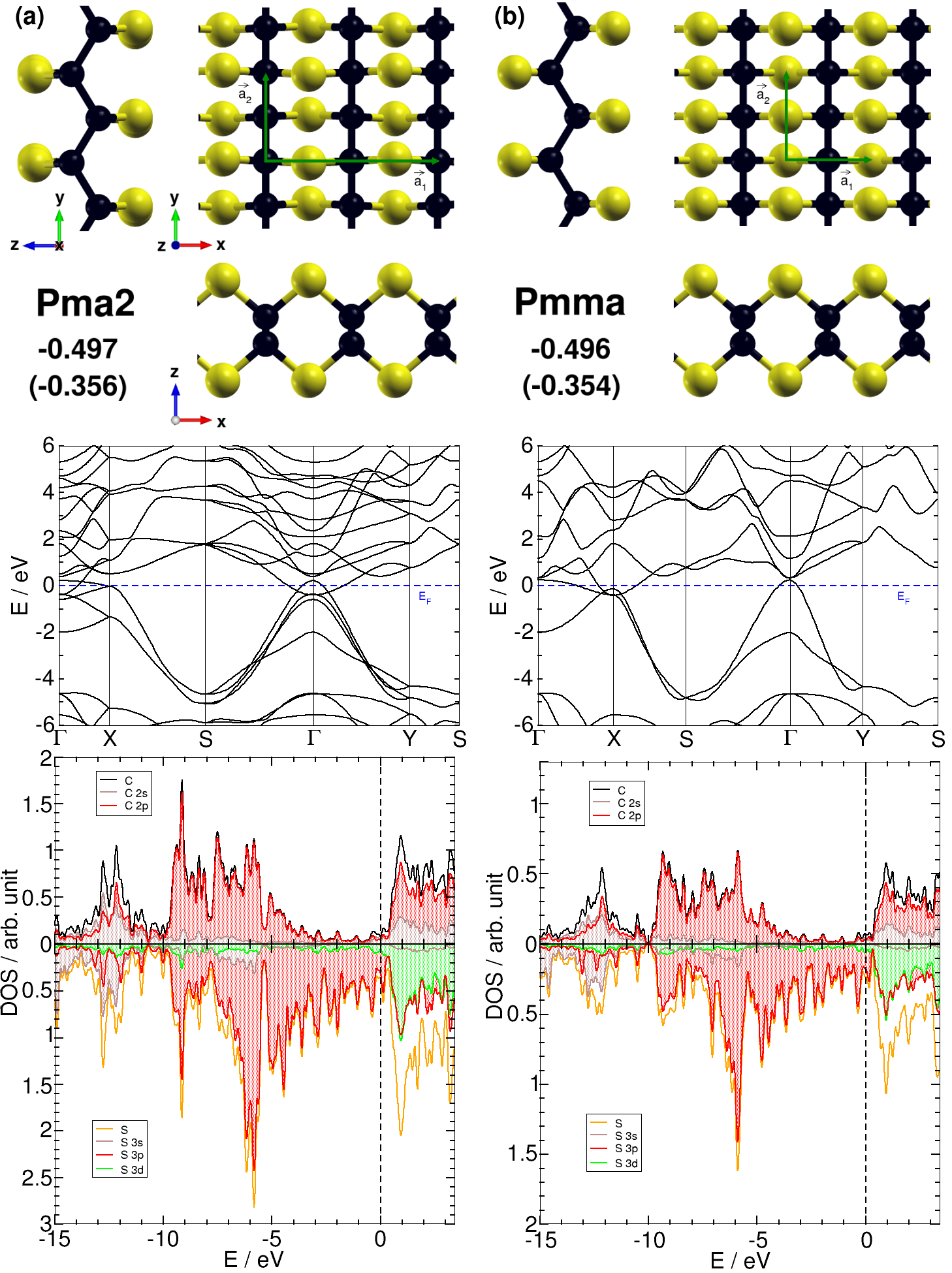}
\caption{\label{thinfilm} 
Geometries for two carbon-sulfur thin films in the (a) \textit{Pma2} and (b) \textit{Pmma} structures. The unit cells employed for the electronic analysis are shown in green. Energies per atom with the SIESTA and VASP methods relative to the $\alpha$ monolayer are given on top and within parenthesis, respectively. Band structures and projected density of states are shown in the lower part.}
\end{figure*}

Carbon group elements form buckled monolayers, such as in silicene and germanene, because $sp^3$ hybridization is favored rather than $sp^2$ hybridization. Silicon nanostructures are expected to gain energy when higher silicon coordination is allowed, as reported for silicon monosulfide thin films\cite{yang2016two,alonso2016array}. 
We study carbon monosulfide following the silicon monosulfide thin films structures.
We find that there is a large energy gain with respect to the $\alpha$ monolayer because a fourth bond is achieved for the carbon having $sp^3$ hybridization. On the one hand, the most stable monolayer do not exhibit a purely carbon $sp^2$ hybridization, but it has a strong $sp^3$ character with a dangling bond. 
On the other hand, sulfur prefers to have two bonds instead of three in order to fulfill the octet rule\cite{yang2016two}. Consequently the energy per atom increases notably with respect to monolayers because the carbon dangling bond is passivated and a sulfur bond is lost.

Figure \ref{thinfilm} shows the detailed geometries of $Pma2$ and $Pmma$ structures. The coordination number, number of bonds, and distances are very similar for both structures. The difference between the $Pma2$ and $Pmma$ structures is the slight distortion around sulfur atoms, which requires considering a large unit cell. The unit cell of $Pma2$ is large than that of $Pmma$, as shown by the unit cell vectors in Fig. \ref{thinfilm}.
 For the $Pma2$ structure,  the sulfur angle between the two C-S bonds is 99$\degree$. Carbon atoms bind to two carbon atoms located at 1.71~\AA{} and two sulfur atoms at 1.86~\AA{}. The angles between bonds established by carbon atoms are all close to 109.5$\degree$, the typical angle for an $sp^3$ hybridization as for methane. The description for the $Pmma$ case reveals almost identical findings.

The band structures of the $Pma2$ and $Pmma$ structures are presented in Fig. \ref{thinfilm}. The electronic structure of these structures look similar as expected. However, the structural distortion of $Pma2$ is splitting the degenerate bands of $Pmma$. The conduction bands close to the Fermi energy expand within a broad range of more than 4 eV. We find that there is no band gap, which means the carbon monosulfide thin films are metallic. The states are metallic in the band crossing the Fermi energy around the $\Gamma$ point and in the $\Gamma X$ direction, regions to be held responsible for electronic conduction.

The projected density of states of the $Pma2$ and $Pmma$ structures shown in Fig. \ref{thinfilm} for both cases are similar, as expected. For carbon we distinguish three zones: an interval [-15,-10] eV with $sp$ hybridization, an interval [-10,-4] with almost all the p-type contribution, and an interval [-4,0] where the contribution from carbon is very small compared to that of sulfur. For sulfur there is large s-type contribution very deep in energy for the interval [-15,-11], and p-type contribution is largely predominant in energies around the Fermi level. In fact, the sulfur p-type states near the Fermi energy are mostly responsible for conduction.

Molecular dynamics simulations show that the two proposed thin films $Pma2$ and $Pmma$ are not stable at room temperature. The structures break into separated parts, showing preference for chain-like geometries in agreement with our results given above. The absence of a band gap for the $Pma2$ and $Pmma$ structures could be related to the low stability of carbon monosulfide thin films.

\subsection{Phonon dispersion and molecular dynamics}

We have carried out molecular dynamics simulations using the Nose-Hoover thermostat in order to study the stability of the carbon monosulfide chains, thin films, and monolayers at room temperature. We show the results for the two most stable monolayers $\alpha$ and $\delta$ and for the \textit{hexa} chains arrays, which are found to be stable at room temperature, in Fig. \ref{dinmol-alpha} and Fig. \ref{dinmol-hexa}. 
Molecular dynamics simulations of the thin film geometries $Pma2$ and $Pmma$ for carbon monosulfide show  that they are not stable at room temperature.

\begin{figure*}[thpb]
      \centering
\includegraphics[clip,width=\textwidth,angle=0,clip]{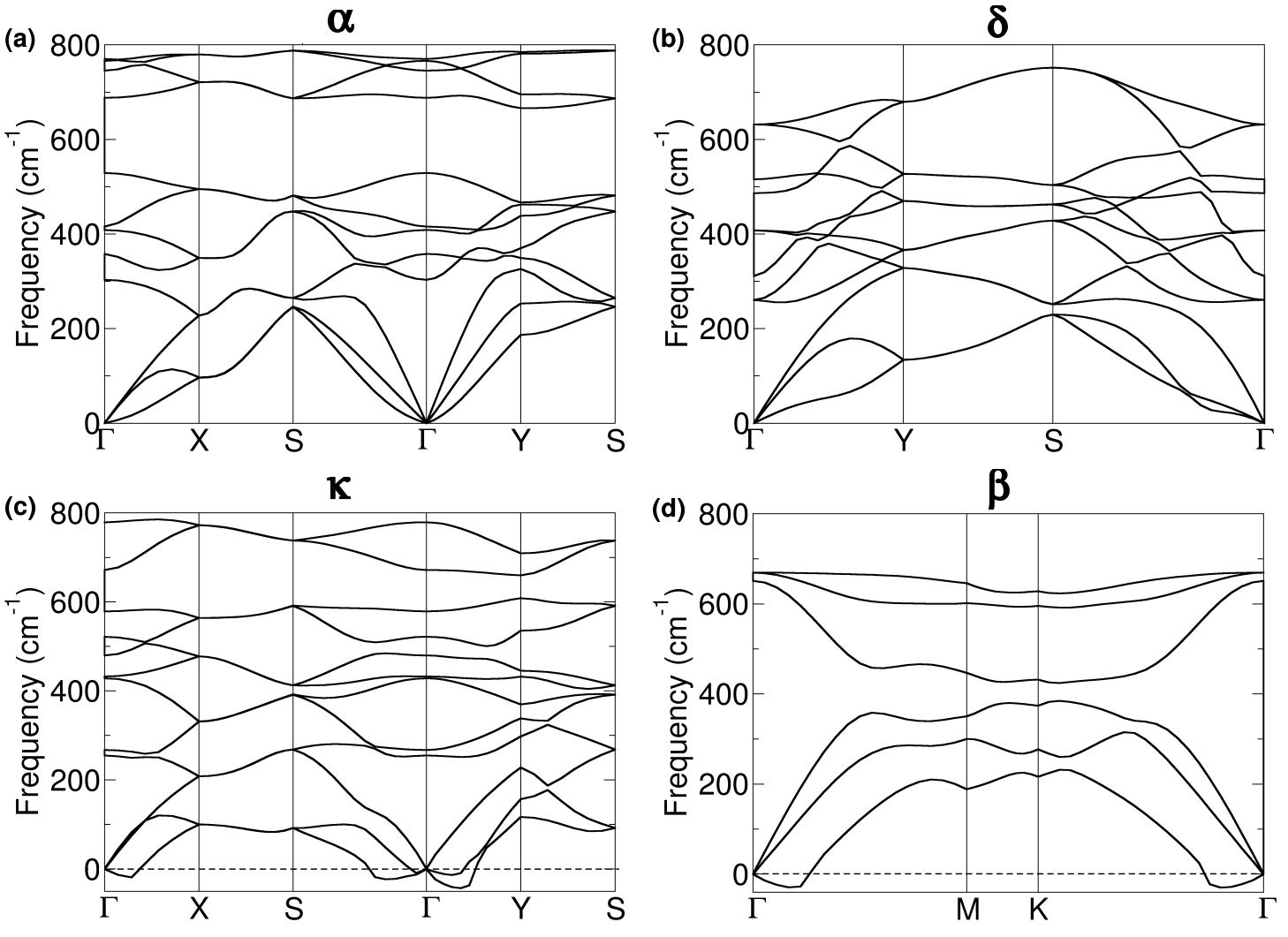}
\caption{\label{figure3}
Phonon dispersions of the (a) $\alpha$, (b) $\delta$ , (c) $\kappa$, and (d) $\beta$ carbon monosulfide monolayers.}
\end{figure*}

\begin{figure*}[thpb]
      \centering
\includegraphics[clip,width=\textwidth,angle=0,clip]{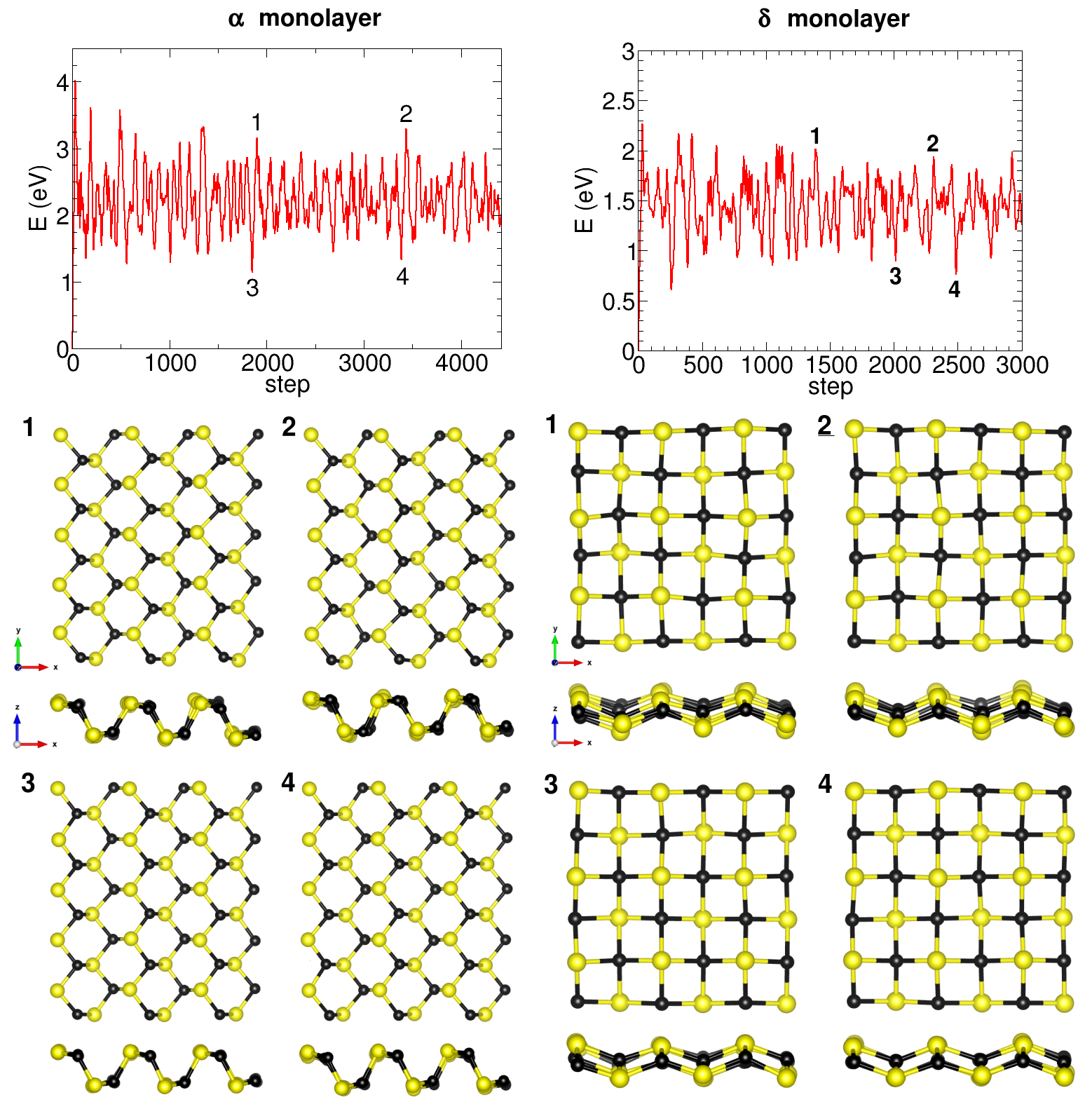}
\caption{\label{dinmol-alpha} 
Energy versus time step for the $\alpha$ (a) and $\delta$ (b) monolayers in a Nose thermostat at 300K. Structural snapshots  are shown below at the four steps marked up in the graph for each case.}
\end{figure*}

\begin{figure}[thpb]
      \centering
\includegraphics[clip,width=0.5\textwidth,angle=0,clip]{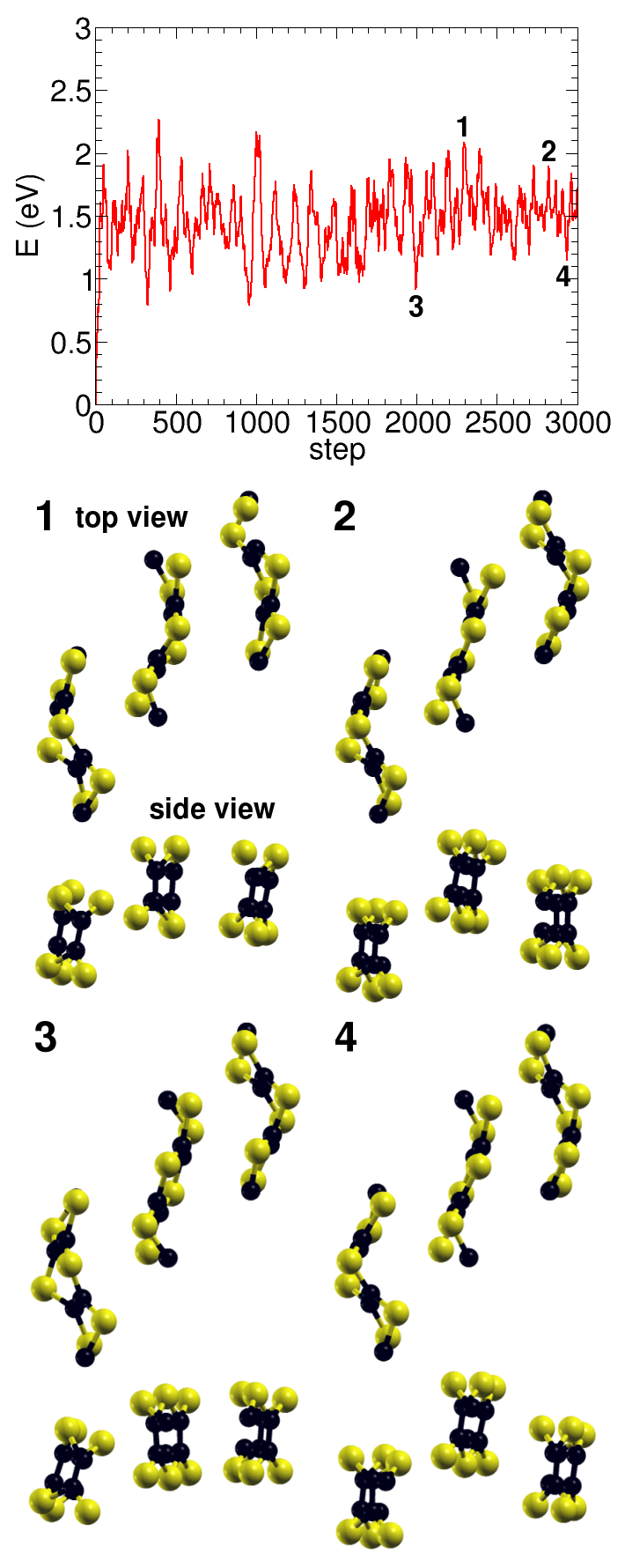}
\caption{\label{dinmol-hexa} 
Energy versus time step for the \textit{hexa} chains in a Nose thermostat at 300K. Structural snapshots are shown below at the four steps marked up in the graph.}
\end{figure}

Phonon dispersions are calculated using linear response in the {\sc Quantum ESPRESSO} \cite{QE-2009,giannozzi1991ab,baroni2001phonons} method, a plane wave code that uses ultrasoft pseudopotentials\cite{pseudo} with a cutoff energy of 50 Ry and the PBE GGA.
We have also computed the phonon dispersion curves for $\alpha$ and $\delta$ carbon monosulfide monolayers, as shown in  Fig. \ref{figure3}. There are no negative frequencies.

\end{document}